\documentclass[fleqn,usenatbib]{mnras}

\usepackage{newtxtext,newtxmath}

\usepackage[T1]{fontenc}

\DeclareRobustCommand{\VAN}[3]{#2}
\let\VANthebibliography\thebibliography
\def\thebibliography{\DeclareRobustCommand{\VAN}[3]{##3}\VANthebibliography}

\usepackage{graphicx}	
\usepackage{xcolor}

\newcommand{\kepler}[0]{\emph{Kepler}}
\newcommand{\gaia}[0]{\emph{Gaia}}
\newcommand{\mmode}[1]{\ifmmode{#1}\else{$#1$}\fi}
\newcommand{\Dnu}[0]{\mmode{\Delta\nu}}
\newcommand{\numax}[0]{\mmode{\nu_\text{max}}}

\newcommand{\Teff}[0]{\mmode{T_\text{eff}}~}

\title[Timing the formation of the Galactic thin disc with asteroseismic stellar ages]{Timing the formation of the Galactic thin disc with asteroseismic stellar ages}

\author[Yaqian Wu]{Yaqian Wu$^{1,2}$\thanks{E-mail: wuyaqian@nao.cas.cn}, Maosheng Xiang$^{1, 3}$, \thanks{E-mail: msxiang@nao.cas.cn},
Gang Zhao$^{1,4}$,\thanks{E-mail: gzhao@nao.cas.cn}, Yuqin Chen$^{1,3,4}$, Shaolan Bi$^{2,3}$,
 \newauthor Yaguang Li$^{5,6}$ \\
$^{1}$CAS Key Laboratory of Optical Astronomy, National Astronomical Observatories, Chinese Academy of Sciences, Beijing 100101, P.\ R.\ China;\\
$^{2}$Department of Astronomy, Beijing Normal University,
             Beijing 100875, P.\ R.\ China;\\
$^{3}$Institute for Frontiers in Astronomy and Astrophysics, Beijing Normal University,  Beijing 102206, P.\ R.\ China;\\
$^{4}$School of Astronomy and Space Science, University of Chinese Academy of Sciences,
             Beijing 101408, P.\ R.\ China;\\
$^{5}$Sydney Institute for Astronomy (SIfA), School of Physics, University of Sydney, NSW 2006, Australia ;\\
$^{6}$Stellar Astrophysics Centre, Department of Physics and Astronomy, Aarhus University, Ny Munkegade 120,
              DK-8000 Aarhus C, Denmark;\\
             }

\date{Accepted XXX. Received YYY; in original form ZZZ}

\pubyear{2015}

\begin{document}
\label{firstpage}
\pagerange{\pageref{firstpage}--\pageref{lastpage}}
\maketitle

\begin{abstract}
The formation of the extended thin disc is the most spectacular event of our Galaxy in the past $\sim8$\,Gyr. To unveil this process, obtaining precise and accurate stellar ages for a large sample of stars is essential although challenging.
In this work, we present the asteroseismic age determination of 5306 red giant branch stars using \kepler{} and LAMOST data, with a thorough examination of how the age determination is affected by the choice of different temperature scales and stellar models. Thanks to the high precision of the asteroseismic and spectroscopic parameters of our sample stars, we are able to achieve age determination with an average accuracy of 12 per cent. However, the age determination is sensitively dependent on the adopted temperature scale, as 50\,K difference in effective temperature may cause larger than 10 per cent systematic uncertainty in the age estimates. Using the ages derived with the most plausible set of the temperature scale, we study the age distribution of the chemical thin disc stars, and present an estimate of the formation epoch of the first Galactic thin disc stars. We find that the first (oldest) thin disc stars have an age of $9.5^{+0.5(\rm rand.)+0.5(\rm sys.)}_{-0.4(\rm rand.)-0.3(\rm sys.)}$\,Gyr, where the systematic uncertainties reflect ages estimated using different stellar evolutionary models. At this epoch, the Galactic thick disc was still forming stars, indicating there is a time window when both the thin and thick discs of our Galaxy were forming stars together. Moreover, we find that the first thin disc stars exhibit a broad distribution of Galactocentric radii, suggesting that the inner and outer thin discs began to form simultaneously.
\end{abstract}




\section{Introduction}
It is well known that the Milky Way's disc is composed of two major components, namely the thick disc and the thin disc. Stars in the thin and thick disc exhibit different distributions in many aspects, such as spatial structure, age, metallicity and kinematics \citep[e.g.][]{Gilmore1983, Juric2008, Rix2013, Hayden2014, Bland2016, Wu2021}. These components reflect two major phases of the Milky Way's assembly history. The thick disc is formed in the first 5\,Gyr of our Galaxy's history, from 13\,Gyr to around 8\,Gyr ago, while the thin disc formation has been the most spectacular event of our Galaxy in the past $\sim8$\,Gyr \citep[e.g.][]{Haywood2013, Snaith2014, Xiang2015, Xiang2022, Sahlholdt2022}.

However, the detailed transition between the thick and thin disc formation still needs to be better understood. On the one hand, chemical evolution models that incorporate a sudden metal-poor gas infall at 7-9\,Gyr ago and an instantaneous mixing with the enriched gas residue of the thick disc can qualitatively reproduce locus of the stellar distribution in the [Fe/H]-[$\alpha$/Fe] plane \citep{Grisoni2017, Spitoni2019, Spitoni2021, Lian2020a, Lian2020b}, suggesting that a strictly sequential transition from the thick disc formation to the thin disc formation is plausible. Simulations in the cosmological framework have also predicted episodic gas infalls, which lead to stellar bimodality in the [Fe/H]-[$\alpha$/Fe] plane and multiple sequences in the age-metallicity plane  \citep[e.g.][]{Calura2009, Wang2022}, both are qualitatively in line with observations. On the other hand, it has been argued that the stellar [Fe/H]-[$\alpha$/Fe] bimodal distribution can be well reproduced with a continuous star formation history when stellar migration effect is considered \citep{Schonrich2009,Buck2020,Sharma2021}.

Observed stellar age distributions of the thin and thick disc populations revealed that there could be a time overlap between the formation of the two discs, as the thick disc stars can reach 8\,Gyr or below at the younger side, while the oldest thin disc stars can be 8-10\,Gyr or even older \citep[e.g.][]{Xiang2017, Wu2018, Wu2019, Silva2021, Xiang2022}. Based on such a time overlap, it has been suggested the thin and thick discs are co-formed \citep{Silva2021,Gent2022}. To reach such a conclusion, one needs accurate stellar age estimates and a robust determination of their uncertainties.

There have been extensive works on the age determination for thin disc stars in the solar neighbourhood, which can give us some hints about when the first thin disc stars were formed. Utilizing nearby F and G type main-sequence, main-sequence turn-off (MSTO) and sub-giant stars, \citet{Fuhrmann1998} and \citet{Bensby2003} found that the age of most thin disc stars is younger than 8--9\,Gyr. \citet{Nordstrom2004} presented the age-metallicity distribution for 462 F and G type stars that have robust age estimates out of 14,000 targets in the Geneva-Copenhagen survey, and their results show that some thin disc stars can be older than 10\,Gyr. Similar results are also presented in \citet{Bergemann2014}. Utilizing 1111 solar-neighbourhood F, G, K stars that have high-resolution spectroscopic abundance determinations, \citet{Haywood2013} suggest that most inner thin disc stars have an age younger than 8\,Gyr, while the outer thin disc stars could have an age of 9--10\,Gyr. Utilizing a large sample of MSTO and subgiant stars from the LAMOST survey, \citet{Xiang2017} found that chemical thin-disc star sequence in the [Fe/H]-[$\alpha$/Fe] plane already became prominent in the age interval of 8--10\,Gyr, and a similar conclusion is reached by \citet{Wu2019} using a large sample of LAMOST red giant branch (RGB) stars with data-driven age estimates based on asteroseismic calibration. With a sample of 11,000 MSTO stars from the H3 survey \citep{Conroy2019}, \citet{Bonaca2020} studied the age distributions of the different Galactic components, and their results show that the low-$\alpha$ thin disc stars can be older than 10\,Gyr. With high-precision age estimates of MSTO and subgiant stars using Gaia parallax, \citet{Sahlholdt2022} and \citet{Xiang2022} suggest that the thin disc become prominent in the age-[Fe/H] plane since 8\,Gyr ago. In summary, these studies suggest that the first thin disc stars are likely to be older than 8\,Gyr.

However, in most cases, uncertainties in the stellar age determination for individual stars are large, with typical precision no better than 20 per cent. Moreover, systematic uncertainty in the stellar age determination is expected to be significant but is usually not well quoted. To uncover the early formation history of the Galactic disc calls for accurate and precise stellar age determination, which only becomes possible recently thanks to the availability of unprecedented data precision from space missions such as \kepler{} \citep{Borucki2010} and \gaia{} \citep{Gaia2016}.

Asteroseismology has proven to be an effective way of precision age-dating method for evolved stars, including sub-giants \citep{Serenelli2017, Wu2017} and red giants \citep[e.g.][]{Silva2018, Wu2018, Pinsonnuault2018, Huang2020, Miglio2021}. This is because the age of a low-mass giant star is dominated by the lifetime of its main-sequence evolutionary phase, which is determined mainly by the stellar mass that can be inferred precisely with asteroseismology.
Combining the asteroseismic parameters from the \kepler{} mission \citep{Borucki2010} and spectroscopic parameters from APOGEE \citep{Majewski2017} and LAMOST \citep{Zhao2012} surveys, \citet{Silva2018, Miglio2021} and \citet{Wu2018} estimated ages for thousands of red giant stars, and their results show a significant number of thin disc stars older than 10\,Gyr. However, the mean age precision of these red giant star samples is $\sim$ 25 per cent, which is still too large to precisely date the formation epoch of the first Galactic thin disc stars.

In this paper, we present improved asteroseismic age determination for 5306 red giant stars observed by \kepler and LAMOST, and characterise the age of the oldest thin disc stars, thus to constrain the formation epoch of the first Galactic thin disc stars. The age precision of the current work reaches 12 per cent, which is a significant improvement compared to earlier work.
This improvement is achieved mainly because of an update of the spectroscopic parameters from the LAMOST spectra. Internal precision in effective temperature and metallicity derived from the LAMOST spectra with DD-Payne can reach 30\,K and 0.05\,dex, separately \citep{Xiang2019}. These values are significantly smaller than those adopted in previous work, which are $\sim100$\,K and $\sim0.1$\,dex \citep{Wu2018}. Furthermore, we carefully characterise systematic uncertainties in the age determination by examining different temperature scales in literature.

The paper is organised as follows.
In Section\,2, we present our data sample and the determination of stellar properties, including \Teff{} and age. In Section\,3, we study the chemistry and kinematics of the sample, and characterise the age of the first thin disc stars. In Section\,4, we discuss when and where the first thin disc stars were formed. Finally, we conclude in Section\,5.

\section{Data}
\subsection{The LAMOST-\kepler{} sample}

The \kepler{} mission \citep{Borucki2010} has provided exquisite data to perform an ensemble of asteroseismic analysis on solar-like stars.
The short-cadence data are primarily designed for asteroseismology study of the main sequence and sub-giant stars, while the long-cadence data are suitable to study oscillations in red giant stars.
In this work, we utilize the global asteroseismic parameters, the frequency of maximum power $\nu_{\rm max}$ and the p-mode large frequency separation $\Delta\nu$, determined by the SYD pipeline \citep{Huber2009} for 16,094 red giants with four years of \emph{Kepler} data \citep{Yu2018}.
The $\nu_{\rm max}$ measurements have a typical uncertainty of 1.6 per cent, and the $\Delta\nu$ measurements have a typical uncertainty of 0.6 per cent.
Furthermore, \citet{Yu2018} also compiles the evolutionary stages (RGB/RC) classified based on asteroseismology \citep{Bedding2011,Stello2013,Mosser2014,Vrard2016,Elsworth2017,Hon2017}.

\begin{figure*}
	\includegraphics[width=\linewidth]{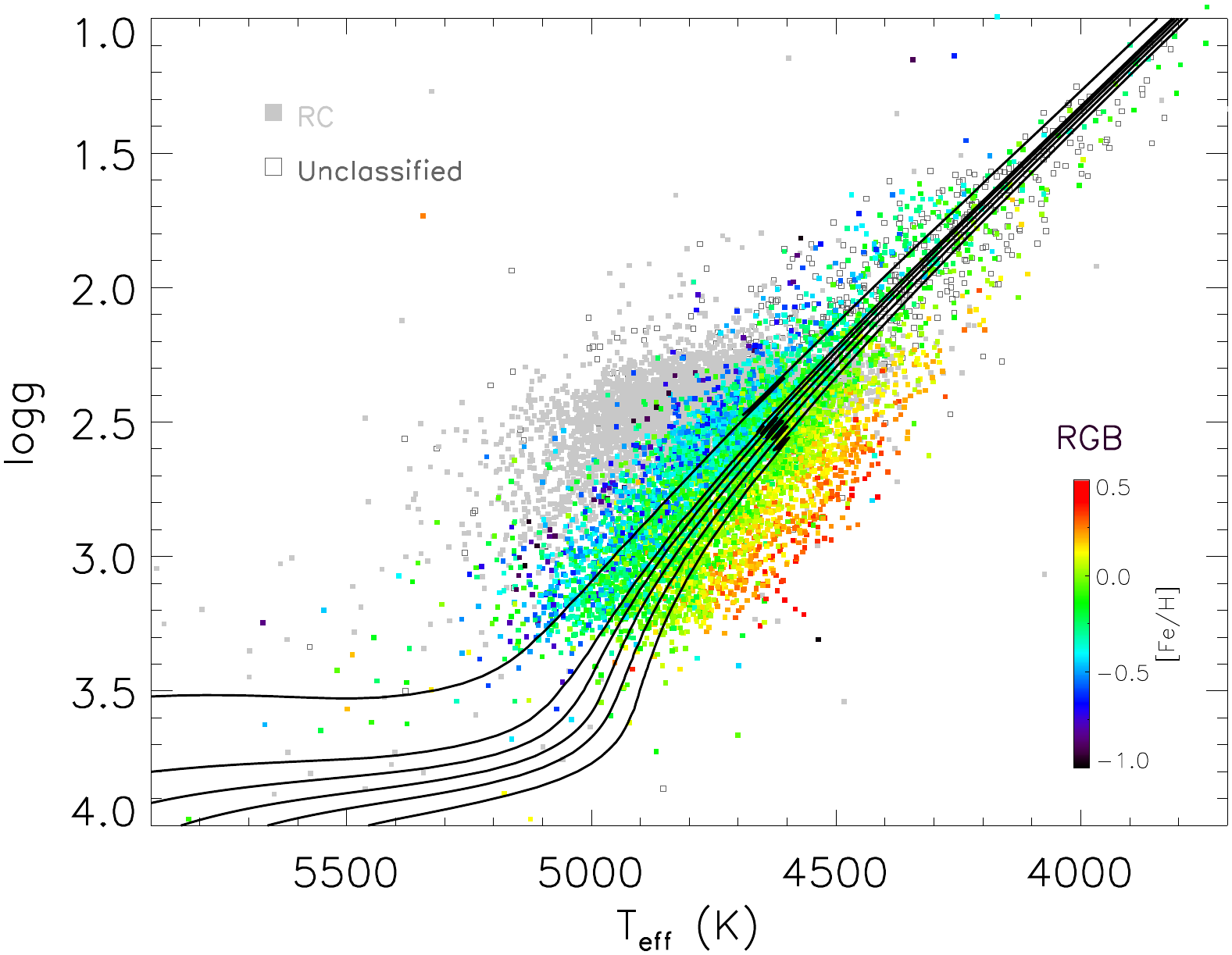}
    \caption{Distribution of stars with \emph{Kepler} asteroseismic parameters and LAMOST spectroscopic parameters in $T_{\rm eff}-\log\,g$ (Kiel) diagram. Stars in different evolutionary phases are shown separately: RGB stars are shown in solid squares, color-coded by their metallicity. Symbols in grey show the RC (filled squares) and unclassified objects (open squares). These classifications are adopted from a complication of literature results \citep{Bedding2011,Stello2013,Mosser2014,Vrard2016,Elsworth2017,Hon2017}. The black solid curves are the Dartmouth Stellar Evolution Database isochrones \citep{Dotter2008} of solar metallicity (${\rm [Fe/H]}=0$), with age of 2,4,6,8,10, and 13\,Gyr, from left to right, respectively. }
    \label{fig1}
\end{figure*}

By June, 2019, the LAMOST-\emph{Kepler} project \citep{De2015} collected more than 200,000 low-resolution ($R\sim1800$) optical spectra ($\lambda$\,3800 -- 9000\,${\AA}$) in the \emph{Kepler} field utilizing the LAMOST spectroscopic survey telescope \citep{Cui2012}. Spectroscopic stellar parameters are delivered from the LAMOST spectra with several pipelines \citep{Luo2015, Xiang2015, Xiang2019, Zhang2020}. In this work, we adopt the LAMOST stellar parameters and abundances catalogue of \citet{Xiang2019}, which contains $T_{\rm eff}$, $\log$\,$g$, and individual elemental abundances for 16 elements (C, N, O, Na, Mg, Al, Si, Ca, Ti, Cr, Mn, Fe, Co, Ni, Cu, Ba) derived with the data-driven Payne (DD-Payne) method \citep{Ting2017,Xiang2019}. For spectra of signal-to-noise ratios (SNRs) higher than 50, the DD-Payne stellar parameters have typical internal precision of about 30\,K for $T_{\rm eff}$, 0.07\,dex for $\log$\,$g$, 0.03--0.1\,dex for elemental abundances, except for Cu (0.5\,dex) and Ba (0.2\,dex).
A cross-identification of LAMOST DR5 and the \citet{Yu2018} \emph{Kepler} asteroseismic sample yields 10,972 common stars, 5306 of which are RGB stars based on the classification in the catalogue of \citet{Yu2018}. Fig.\ref{fig1} shows the distribution of the sample stars in the Teff-logg (Kiel) diagram. In the figure, we also show the RCs and unclassified stars.

\subsection{Calibration of effective temperature}
\begin{figure*}
    \includegraphics[width=\linewidth]{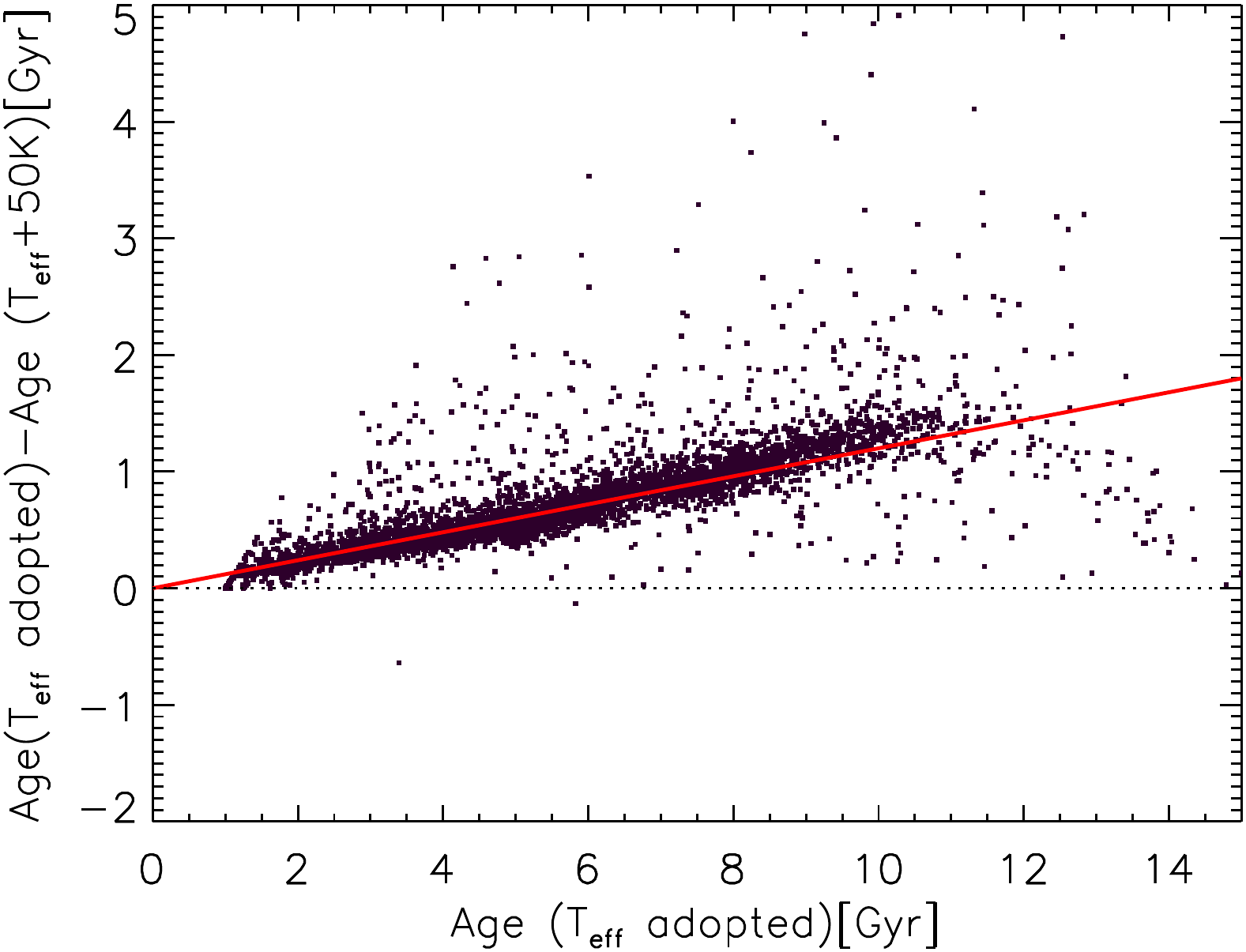}
    \caption{Differential ages between estimates using effective temperature adopted by the current work and estimates after adding a 50\,K to the adopted temperature, as a function of the former. The red lines represents a constant relative age difference of 12 per cent. It illustrates that a 50\,K offset in effective temperature may cause a systematic error in the age estimates by more than 10 per cent. }

    \label{fig:4}
\end{figure*}

In our method of age determination (Sect.\,2.3), the effective temperature works in two places: first, it is directly used as an observable to constrain the probability distribution function of the age, and second, it is used along with the asteroseismic frequencies to infer the stellar mass via the asteroseismic scaling relation. Therefore, accurate age determination of RGB stars relies on accurate effective temperature estimates. This is especially important considering the fact that stellar isochrones of different ages spread in a small effective temperature range in the RGB phase, so any small systematic errors in effective temperature may cause a large deviation in the age determination. As an illustration, Figure~\ref{fig:4} shows that a 50\,K systematic difference in effective temperature may cause more than 10 per cent systematic error in the resulting age estimates.

The effective temperature of the DD-Payne catalogue is trained on the APOGEE estimates derived with the \emph{Payne} \citep{Ting2019}, after some corrections in order to match better with the stellar isochrones. For RGB stars, the PARSEC isochrones \citep{Bressan2012} are adopted to correct for the effective temperature of the training stars adopted by the DD-Payne (see Appendix of \cite{Xiang2019} for details). On top of that, no further external calibrations are implemented to the DD-Payne estimates.

To ensure high accuracy of the effective temperature, in this work, we calibrate the DD-Payne effective temperature to external, well-known photometric temperature scales. We choose to apply two temperature scales, the scale of \citet{Gonz2009} and the scale of \citet{Huang2015}. The scale of \citet{Gonz2009} is based on the infrared flux method \citep[IRFM;][]{Blackwell1977}, and is adopted as the calibration scale of the APOGEE survey \citep{Jonsson2020}. The scale of \citet{Huang2015} is built on direct measurement of the stellar angular size via interferometric data. We will compare the results from these two scales.

To implement the calibration, we derive the photometric effective temperature of our sample stars using the $V-K_{\rm s}$ colour, where the $V$-band photometry is from the APASS survey \citep{Munari2014}, the $K_{\rm s}$-band photometry is from the 2MASS survey \citep{Skrutskie2006}.
The $V-K_{\rm s}$ colour is de-reddened using reddening $E(B-V)$ deduced from the star-pair method, which has been extensively used to deduce $E(B-V)$ for stars with spectroscopic parameters \citep{Yuan2013, Yuan2015, Xiang2017, Xiang2019}. The idea is that given the availability of spectroscopic parameters ($T_{\rm eff}$, log\,$g$, [Fe/${\rm H}$]), the intrinsic colours of the stars can be well determined by using stars in a control field that has well-known extinction, specifically, stars at Galactic latitudes that the \citet{Schlegel1998} extinction map is accurate \citep{Yuan2013}. This is a data-driven approach, so that the results are not sensitive to systematic errors in the spectroscopic parameters. To transfer $E(B-V)$ to $E(V-Ks)$, we adopt a total-to-selective extinction coefficient of 3.1 in $V$ band, and 0.34 in $K_{\rm s}$ band. For a calibration star, we require the photometric uncertainty to be smaller than 0.05\,mag in both $V$ and $K_{\rm s}$ bands.
Ultimately, we have 1449 stars in the calibration sample.

\begin{figure*}
    \includegraphics[width=\linewidth]{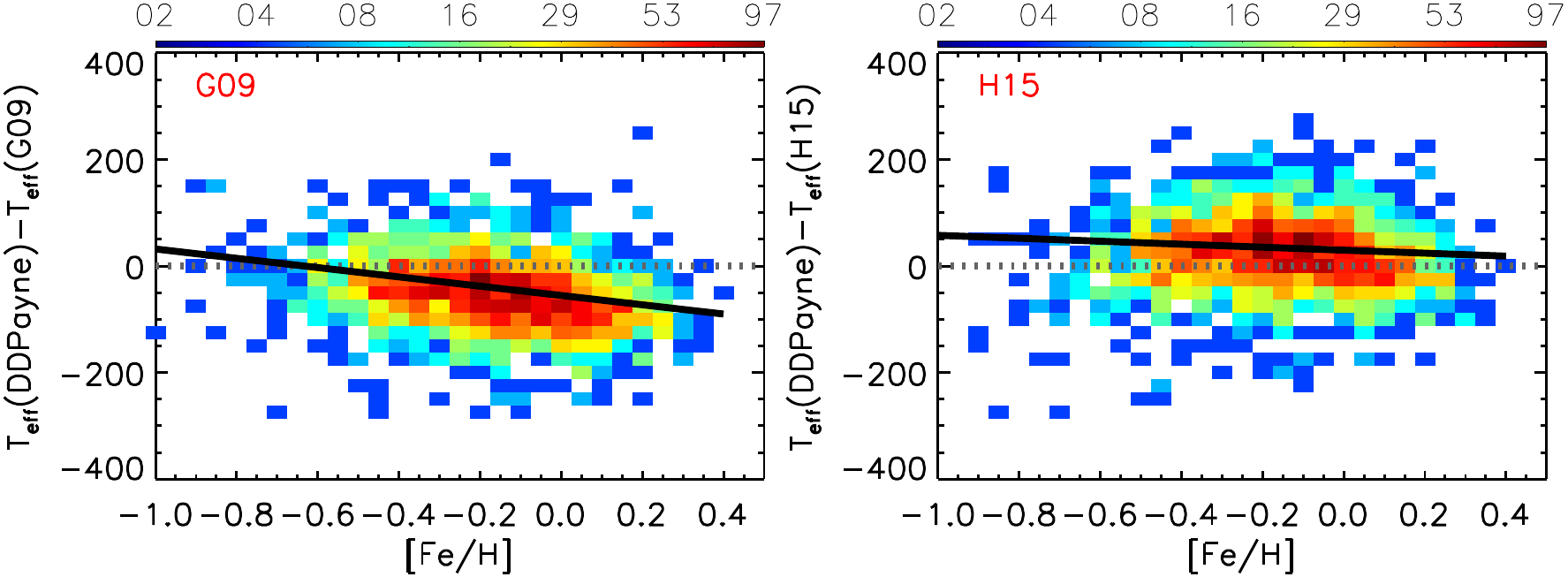}
    \caption{{\em left}: Differential temperature between the LAMOST DD-Payne estimates and the photometric estimates using the \citet{Gonz2009} temperature scale as a function of [Fe/H] with colour-coded stellar number density.
    The solid line in black shows a linear fit to the difference for the disc stars, and it yields a negative trend of $\Delta$Teff= -54.9690 - 86.8584$\times$[Fe/H]. {\em Right}: Same as the left panel, but for temperature difference between the LAMOST DD-Payne estimates and the photometric estimates using the \citet{Huang2015} temperature scale. A linear fit (black) is found to well reproduce the systematic trend for the disc stars: $\Delta$Teff= 29.8236 - 27.8769$\times$[Fe/H].}

   \label{fig:fig2}
\end{figure*}

The left panel of Fig.\,\ref{fig:fig2} presents the difference between the LAMOST DD-Payne and photometric temperature, and it shows a clear trend with metallicity. The DD-Payne temperature is lower than the \citet{Gonz2009} temperature scale by 50\,K at ${\rm [Fe/H]}=0$, and the difference increases to 80\,K at ${\rm [Fe/H]}=0.3$, but decreases to 0 at ${\rm [Fe/H]}=-0.6$. On the contrary, the right panel of Fig.\,\ref{fig:fig2} shows a more or less flat trend with metallicity. The DD-Payne temperature is higher than the \citet{Huang2015} temperature scale by 30\,K at ${\rm [Fe/H]}=0$, and the difference increases to 50\,K at ${\rm [Fe/H]}=-0.2$. We note that the large star-to-star scatter in Fig.\,\ref{fig:fig2} is caused by the stochastic temperature errors of individual stars. However, the mean offsets for both panels are significant.

Finally, for calibrating the LAMOST DD-Payne temperature to the \citet{Gonz2009} scale, we adopt the following linear correction,
\begin{equation*}
    T_{\rm eff} = T_{\rm eff (DD-Payne)} + 54.9690 + 86.7584\times[\rm Fe/H].
\end{equation*}

For calibrating the LAMOST DD-Payne temperature to the \citet{Huang2015} scale, we adopt the correction below,
\begin{equation*}
    T_{\rm eff} = T_{\rm eff (DD-Payne)} - 29.8236 + 27.8769\times[\rm Fe/H].
\end{equation*}

In the following analysis, we will present the results using the effective temperatures calibrated to the \citet{Gonz2009} scale, while leaving those using the \citet{Huang2015} scale in the Appendix. The main consideration here is for self-consistency: the \citet{Gonz2009} scale is built on synthetic model spectra, and the temperature in the isochrones adopted for age determination is also dependent on the stellar and atmospheric models.

\subsection{Age determination}

The age of a low-mass red giant star is determined mainly by the lifetime of its main-sequence evolutionary phase, which depends on the initial mass given the chemical composition is known. Therefore, one can obtain a good age estimate from stellar evolutionary tracks if the stellar mass is known accurately \citep{Ness2016, Martig2016, Wu2018}. With the global asteroseismic frequencies and spectroscopic effective temperature, we estimate the stellar mass utilizing the asteroseismic scaling relation \citep{Ulrich1986,Brown1991,Kjeldsen1995,Kallinger2010}:
 \begin{equation}\label{3}
    \frac{M}{M_{\odot}}=\left(\frac{\nu_{\rm max}}{\nu_{\rm max,\odot}}\right)^{3}\left(\frac{\Delta\nu}{f_{\Delta\nu}\Delta\nu_{\odot}}\right)^{-4}\left(\frac{T_{\rm eff}}{T_{\rm eff,\odot}}\right)^{1.5},
  \end{equation}
where the factor $f_{\Delta\nu}$ is introduced as an empirical correction to the canonical scaling relation, which has been suggested to suffer systematic errors \citep{White2011,Sharma2016,Guggenberger2016,Viani2017,Li2022}. Here we correct for the scaling relation following \citet{Sharma2016}.
We adopt the solar values $T_{\rm eff,\odot}$ = 5777\,${\rm K}$, $\nu_{\rm max,\odot}$ = 3090 $\mu$Hz and  $\Delta\nu_{\odot}$ = 135.1 $\mu$Hz \citep{Huber2011}.
Uncertainty of the resulting mass is estimated by propagating uncertainties of the \numax, \Dnu~ and \Teff with Monte Carlo Markov Chain (MCMC) method.

With the asteroseismic mass, spectroscopic parameters $T_{\rm eff}$, log\,g, [Fe/${\rm H}$], and [$\alpha$/Fe], we infer the stellar age by fitting the stellar isochrones with a Bayesian approach, similar to our previous work \citep{Xiang2017, Wu2018}. We adopt the Dartmouth Stellar Evolution Database stellar isochrones \citep[DESP;][]{Dotter2008}, while the Yonsei-Yale isochrones \citep[$Y^2$;][]{Demarque2004} and the PAdova and TRieste Stellar Evolution Code \citep[PARSEC;][]{Bressan2012} are also used for examination of possible model dependence in the age determination (see Appendix B).
Note that the DD-Payne adopts the solar abundance scale of \citet{Asplund2009}, which is the same as the PARSEC isochrones, while the $Y^{2}$ isochrones adopt the solar abundance scale of \citet{GS96}, and the DSEP isochrones adopt the scale of \citet{GS98}. This means that the definitions of the solar metallicity for DSEP and $Y^{2}$ are higher about 0.1\,dex compared to LAMOST DD-Payne. To mitigate this discrepancy, we decrease the metallicity of the $Y^{2}$ and DSEP isochrones by 0.1\,dex.

\begin{figure*}
    \includegraphics[width=\linewidth]{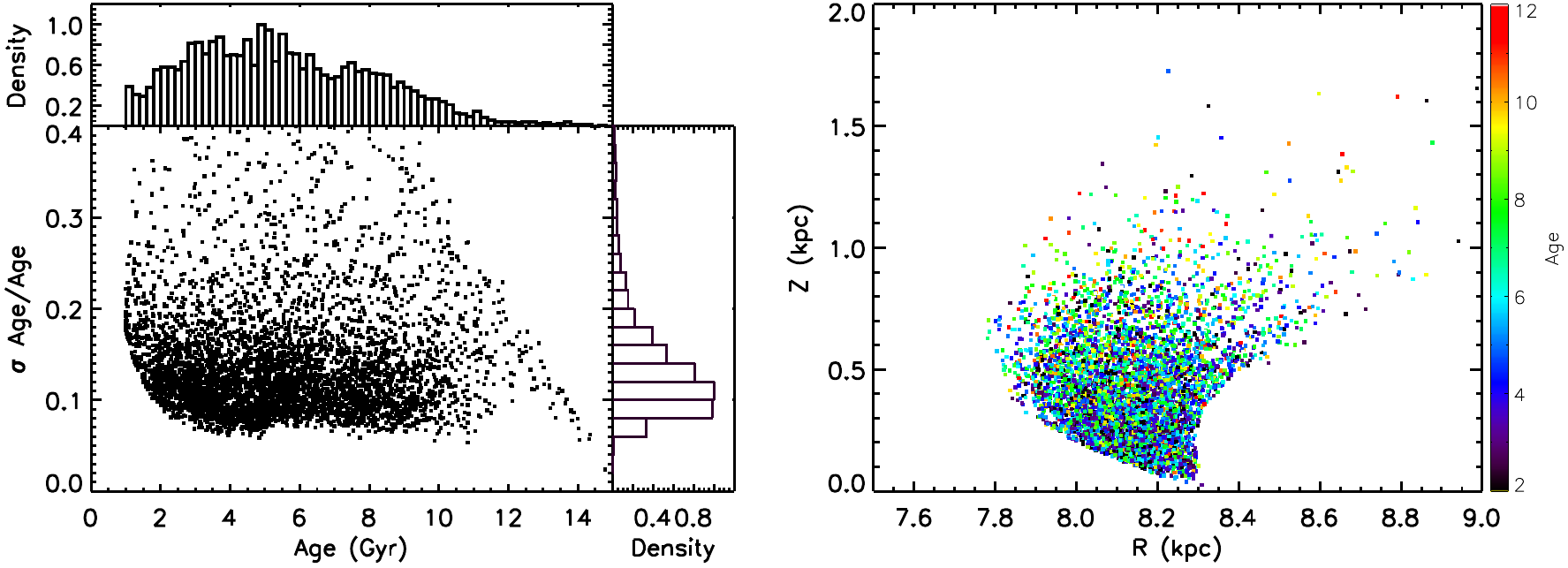}
    \caption{{\em Left}: Distributions of age estimates, as well as their error estimates, for the RGB sample stars. {\em Right}: Spatial distribution of the RGB sample stars with colour-coded ages in the $R$ and $Z$ plane.}
    \label{fig4}
\end{figure*}

The left panel of Fig.~\ref{fig4} shows the distributions of relative age uncertainties for our RGB sample stars.
The sample covers an age range from $\sim1$\,Gyr to the age of the universe ($\sim$\,13.8 Gyr; \citet{Planck2016}).
The median of the relative age uncertainties is about 12 per cent. A small fraction ($\sim$\,8 per cent) of stars exhibit large age errors ($>$\,30 per cent). This is mainly due to large uncertainties in mass estimates and/or in stellar atmospheric parameters.
To our knowledge, this is currently the largest RGB star sample with state-of-the-art asteroseismic age estimates. Compared to \citet{Wu2018}, the age precision of the sample stars has been improved because of both smaller internal (random) uncertainty (30\,K versus 100\,K) and a better characterization of systematic uncertainty in the effective temperature estimates. Compared to other asteroseismic data sets, such as the APOKASC \citep{Pinsonnuault2018}, our data set has a larger number of stars (5306 versus 3623). Notably, as we have adopted the same temperature scale as for the APOGEE, our data set offers a good complementary to the existed APOKASC sample.
The right panel of Fig.~\ref{fig4} shows the spatial coverage of the sample stars in the disc $R$-$Z$ plane. The sample stars are within 7.8\,kpc$<R<8.6$\,kpc and 0\,kpc$<Z<1.5$\,kpc, presumably dominated by thin disc stars, but might also contain a moderate number of thick disc stars. We expect more old stars in the higher $Z$ than in the lower $Z$ because the higher has more thick disc stars than the lower.

\section{Results}
In this section, we select the thin disc stars from our RGB sample based on the locus in the [Fe/H]-[$\alpha$/Fe] plane, and characterise their chemistry, kinematics, and age distribution.
\subsection{Selection of thin disc stars}

\begin{figure*}
    \includegraphics[width=\linewidth]{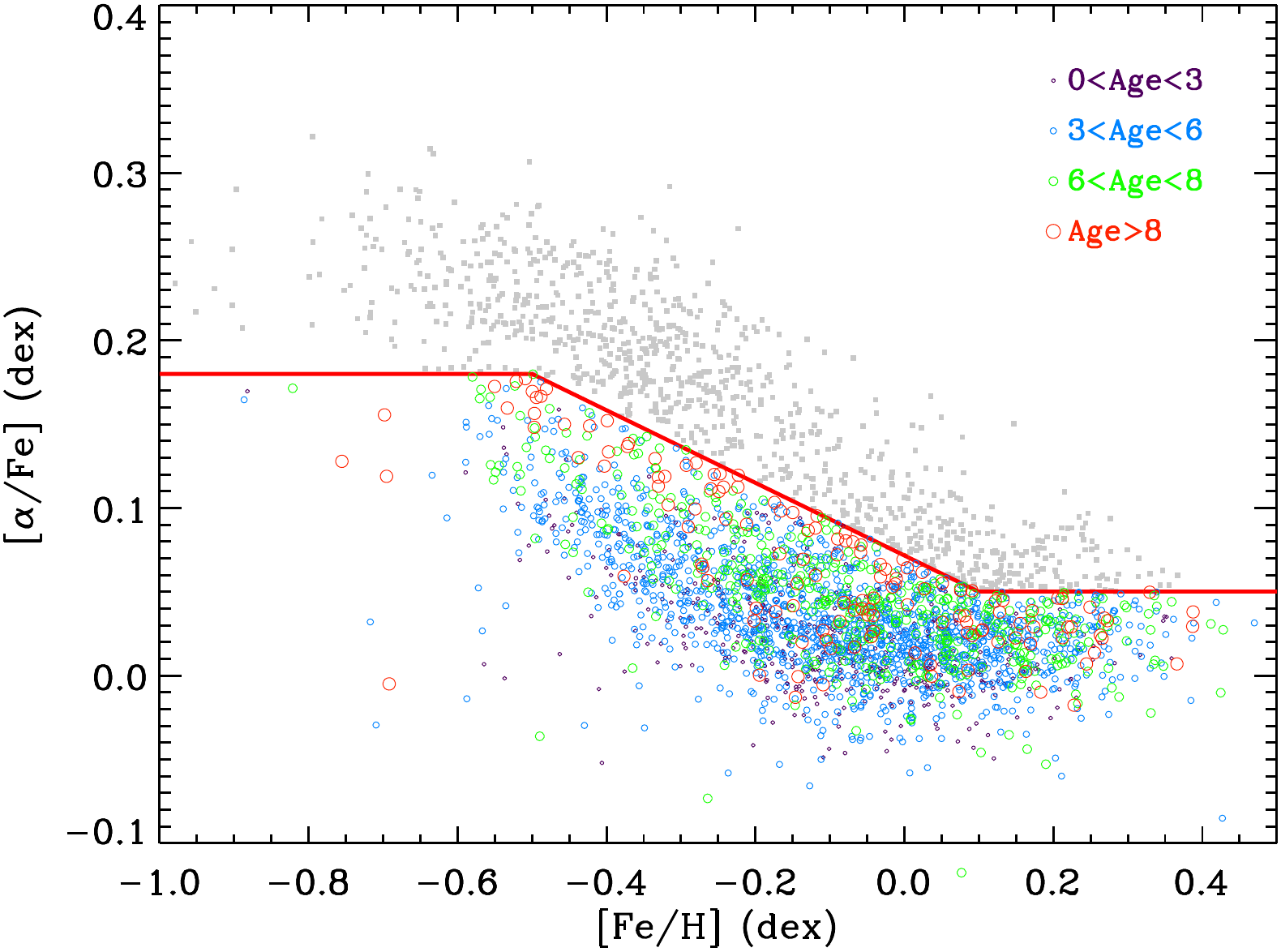}
    \caption{Distributions of the RGB sample stars in the  [$\alpha$/Fe]–[Fe/${\rm H}$] plane. The red lines delineate the demarcation of the chemical thin and thick disc sequences. The thin disk stars are color-coded by their ages: purple circle represent thin disc stars with ages smaller 3\,Gyr, cyan circle represent thin disc stars with ages between 3 to 6\,Gyr, green circle represent thin disc stars with ages between 6 to 8\,Gyr, and red circle represent thin disc stars with ages older than 8\,Gyr. The sizes of the circles are proportional to the stellar ages. Thick disc stars are shown in grey colors.}
    \label{fig:6}
\end{figure*}

It has been well known that the disc stars distribute along two separate sequences in the [Fe/${\rm H}$] -- [$\alpha$/Fe] plane \citep{Fuhrmann1998, Bensby2003, Haywood2013, Hayden2015}. Such a double-sequence feature has also been revealed by the LAMOST RGB sample \citep[e.g.][see also Fig.~\ref{fig:6}]{Wu2018, Zhang2021}. Nevertheless, there is no uniform criterion for separating the chemical thin and thick disc sequences, as different abundance data sets may suffer systematics. We therefore adopt an empirical criterion to select the chemical thin disc stars (Fig.~\ref{fig:6}). Our criterion is expressed as
\begin{equation}
  \begin{cases}
    {\rm [\alpha/Fe]}<0.18, {\rm if [Fe/H]} < -0.5 \\
    {\rm [\alpha/Fe]}<-0.217\times{\rm [Fe/H]}+0.0717, {\rm if} -0.5<{\rm [Fe/H]}< 0.1, \\
    {\rm [\alpha/Fe]} < 0.05,  {\rm if [Fe/H]}>0.1 \\
  \end{cases}
\end{equation}

Furthermore, we discard all stars with $\log\,g<2$ from our sample, as their age estimates are found to suffer large systematic uncertainties and thus unrealistically large compared to stars with $\log\,g>2$. The reason is unclear, but possibly related to some unknown systematic errors in the temperature or asteroseismic mass estimates.

With these selections, we obtain 2537 chemical thin disc stars with relative age errors smaller than 15 per cent. We note that for stars with super solar metallicity (${\rm [Fe/H]}>0$), the thin and thick sequences are less discernible. We expect that the thick disc sequence, if reaches such high metallicity and exhibit higher [$\alpha$/Fe] \footnote{Recent work has suggested that the thick disc sequence has been enriched up to an [Fe/H] of $\sim$0.4\,dex \citep{Xiang2022}}, can still be statistically separated from the thin disc sequence given the robust [$\alpha$/Fe] estimates.

\subsection{Chemistry}
\begin{figure*}
    \includegraphics[width=\linewidth]{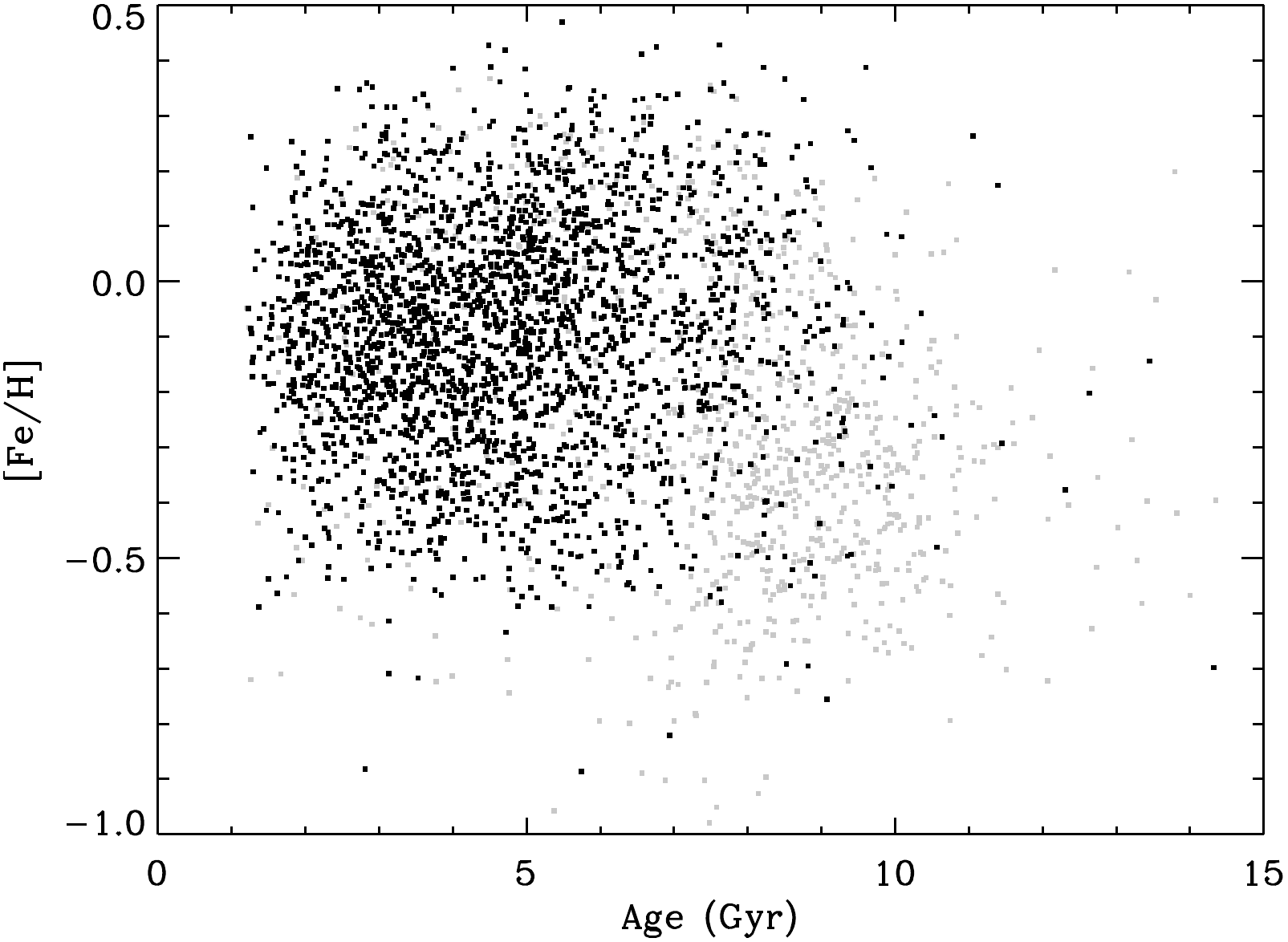}
    \caption{Stellar age-[Fe/${\rm H}$] relation. The black dots represent the chemical thin disc stars, which are the main focus of the current work, and the gray dots represent the chemical thick disc stars.}
    \label{fig:7}
\end{figure*}

The age-[Fe/${\rm H}$] relation of the thin disc sample stars is presented in Fig.~\ref{fig:7}, which shows a broad [Fe/${\rm H}$] distribution at all ages from 1 to $\sim9$\,Gyr. Such a flat relation is consistent with previous findings for solar neighbourhood stars \citep[e.g.][]{Nordstrom2004, Bergemann2014, Xiang2017, Wu2018}. Given the small space coverage of the sample stars (7.8\,kpc $< R <$ 8.6\,kpc), we do not expect that in-situ stars at any constant age can have such a broad [Fe/${\rm H}$] distribution. The broad [Fe/${\rm H}$] distribution is, however, most likely a consequence of stellar radial migration, which mixes stars born at different Galactocentric radii with different [Fe/${\rm H}$] \citep{Roskar2008, Schonrich2009, Loebman2011, Hayden2015, Frankel2018, Sharma2021}.

Other effects, for instance, the sample selection effects, have been suggested to have an impact on the observed age–[Fe/${\rm H}$] relation \citep{Holmberg2009, Marsakov2011, Calura2012}. This is particularly true considering that our sample stars are limited in a small volume, but stars in different parts or components of the Milky Way may exhibit different age-metallicity relation. With larger samples of main-sequence turn-off and subgiant stars, \citet{Sahlholdt2022} and \citet{Xiang2022} found that the stellar age-[Fe/H] plane exhibits complex structures, such as multiple sequences that correspond to different Galactic components. Our sample size is too small to reveal such fine structures. However, within the volume that our sample covered, we do not expect strong selection effects given the uniform target selection strategy of the LAMOST survey \citep[e.g.][]{Liu2014, Yuan2015, De2015, Chen2018}. The impact of the LAMOST selection function on the stellar metallicity distribution has been studied in several works \citep[e.g.][]{Xiang2015, Nandakumar2017, Chen2018, Wang2019}, and no strong effects are found on the metallicity distribution of stars at any given age.

 Interestingly, Fig.~\ref{fig:7} exhibits a sharp cut off of the stellar density at $\sim$9\,Gyr (Sect. 3.4), but even at this age border, the thin disc stars spread a broad [Fe/${\rm H}$] distribution from $-0.6$ to 0.4. This is also illustrated in Fig.~\ref{fig:6}, which shows that the oldest thin disc stars ($\tau>8$\,Gyr) are uniformly distributed in the [Fe/${\rm H}$]--[$\alpha$/Fe] plane. This implies that the oldest thin disc stars might have been born at a broad range of Galactocentric radii (see Sect. 4.2 for discussion).

Fig.~\ref{fig:7} also shows that most of the thick disc stars in our sample have ages of 8 to 10\,Gyr.
 This is younger than previous estimates, which suggest an age of about 11\,Gyr for thick disc stars with ${\rm [Fe/H]}\simeq-0.5$ \citep[e.g.][]{Xiang2022}. We suspect this is because the effective temperatures of the LAMOST DD-Payne for high-$\alpha$ thick disc stars have been overestimated, as a consequence of the fact that the training sets are calibrated to the PARSEC isochrones of {\em solar-scaled} metallicity \citep{Xiang2019}. We therefore choose to focus on the age of the thin disc stars, while leaving the study of the thick disc stars to future work.

\begin{figure*}
   \includegraphics[width=\linewidth]{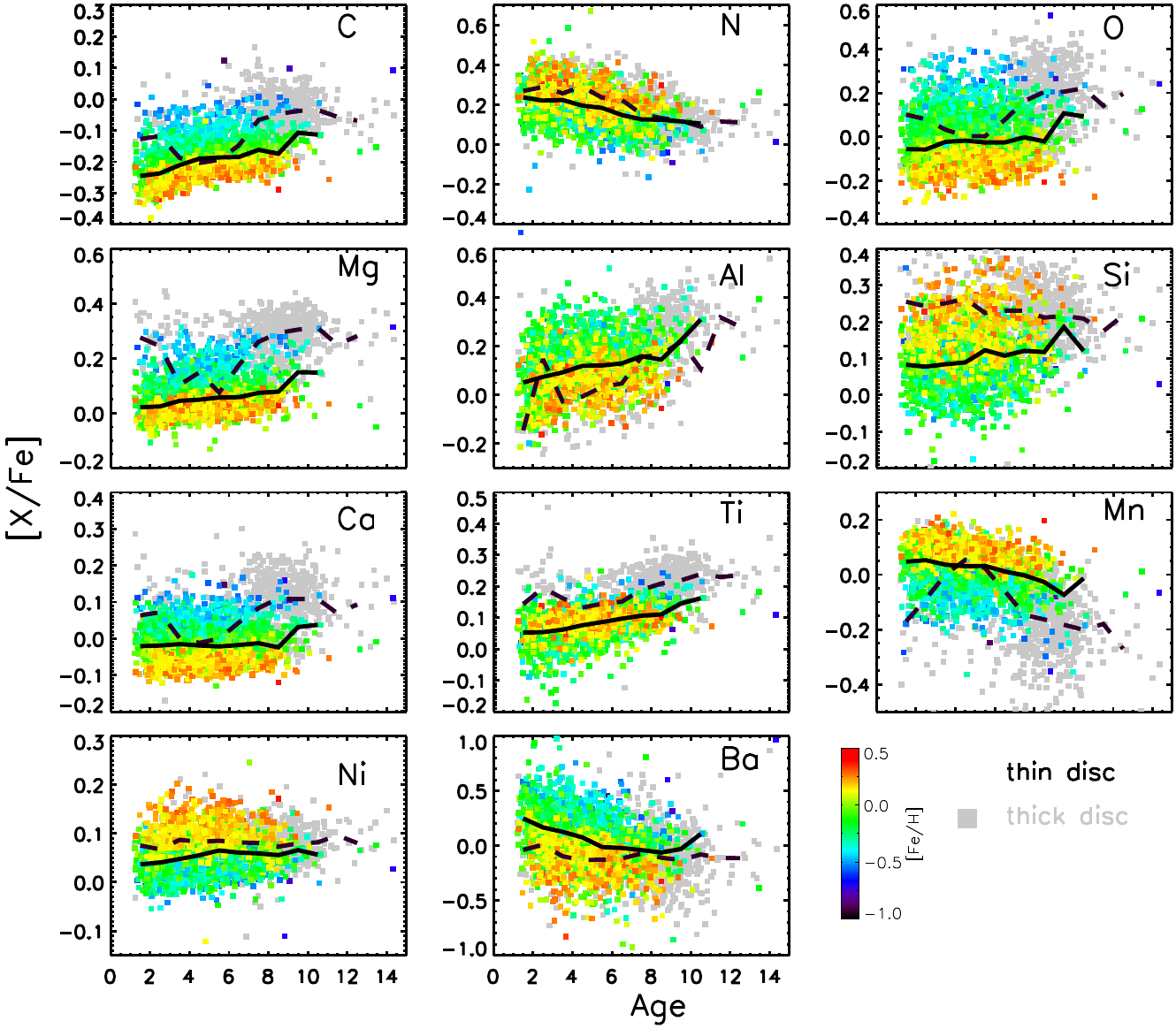}
    \caption{Relation between age and [X/Fe] elemental abundance ratios for the chemical thin disc stars, color-coded by their ${\rm [Fe/H]}$ values, and the thick disc stars (grey). The solid and dashed curves show the median [X/Fe] as a function age for the thin and thick disc, respectively. }

    \label{fig:10}
\end{figure*}

Fig.~\ref{fig:10} presents the abundance ratios [X/Fe] of the sample stars as a function of age for 11 elements, namely, C, N, O, Mg, Al, Si, Ca, Ti, Mn, Ni, and Ba. These elements belong to different nucleosynthetic families, including O, Mg, Si, Ti, Ca, for $\alpha$-elements, C, N for light proton elements and Al for odd-light elements, Mn and Ni for iron peak elements, and Ba for (slow) neutron capture process.
First of all, the figure exhibits a clear difference in the trend of the mean [X/Fe] as a function of age between the thin and thick discs for C, O, Mg, Si, Ca, Ti, Ni, and Mn, validating our criteria to distinguish the thin and thick disc stars.

Secondly, the [X/Fe] for thin disc stars at a given age is sensitively dependent on the  ${\rm [Fe/H]}$. For C, O, Mg, Al, Ca, and Ba, the values of [X/Fe] are higher for more metal-poor stars, while for N, Mn, and Ni, the values of [X/Fe] are lower for more metal-poor stars.
A definitive explanation of these trends requires quantitative comparison with chemical evolution models \citep[e.g.][]{Kobayashi2020}. However, qualitatively, the results are in line with the fact that for more metal-poor stars, a larger fraction of their abundances are enriched by the exposition of type II supernovae. The Mg is purely produced by type II supernovae, so that we observe a clear trend that older or metal-poor stars have higher [Mg/Fe] values. The Mn is almost exclusively produced by type Ia supernovae, which has a long formation time scale, so that the younger stars exhibit higher [Mn/Fe] values than the older stars.

However, we note that the trend for [Si/Fe] is unexpected: the figure shows that the more metal-rich stars have higher [Si/Fe], which is different to the cases of other $\alpha$-elements, such as O, Mg, Ca. This is possibly because of systematic errors in the [Si/Fe] measurements. The DD-Payne [Si/Fe] for giant stars exhibit a positive trend with [Fe/H] at the metal-rich end (see Fig. 18 of \cite{Xiang2019}), which is not seen for dwarf stars (see Fig. 17 of \cite{Xiang2019}). A similar plot from the APOGEE survey is drawn by \citet[][see their Fig.7]{Ness2019}. Among the elements in common, our results of N, O, Mg, Al, Ca, Ti, and Mn are consistent with \citet{Ness2019}. While the results for C, Si, and Ni exhibit some different trends. For [C/Fe], our results show a clear dependence on [Fe/H], while \citet{Ness2019} did not exhibit such a trend. For [Ni/Fe], our results exhibit an opposite trend with \citet{Ness2019}. These differences reflect possible systematic in different survey data sets \citep[e.g.][]{Xiang2019}. Note that at the younger age end ($<6$\,Gyr), the dashed line shows an increase of [X/Fe] with decreasing age, this is an artefact due to the presence of binary evolution products, e.g. \citet{Zhang2021}.

\subsection{Kinematics}
\begin{figure*}
    \includegraphics[width=\linewidth]{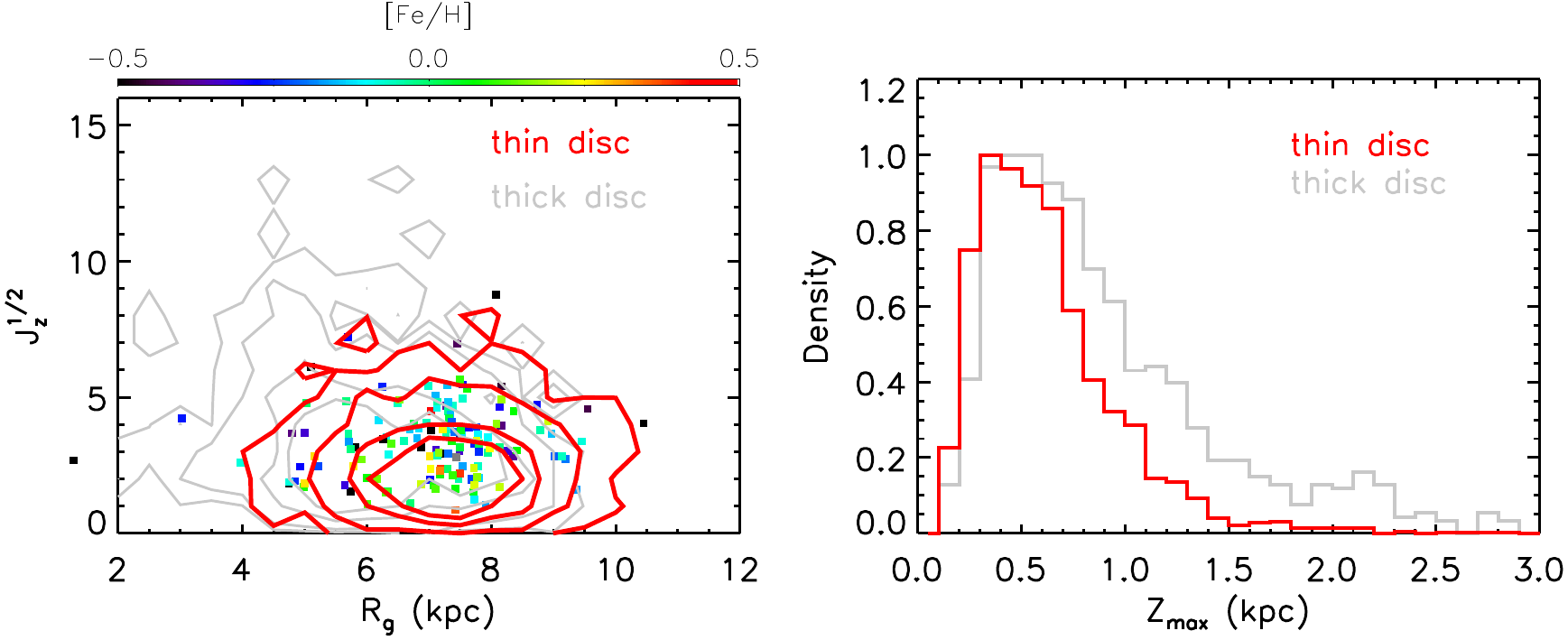}
   \caption{{\em Left}: Distribution contours in the $R_{\rm g}$ and $\sqrt{J_{\rm Z}}$ plane for the chemical thin disc (red) and chemical thick disc (grey) stars. The contours indicates 50,70,90, and 99 per cent of the thin and thick disc stars. Thin disc stars older than 8\,Gyr are shown in squares, color-coded by their metallicity. {\em Right}: Normalized histogram distributions of orbital maximal height $Z_{\rm max}$ for chemical thin disc (red) and thick disc (grey) stars. The histograms are normalized to their peak values. Both panels show significant overlaps between the chemical thin and thick disc stars. }
    \label{fig:8}
\end{figure*}

In Fig.~\ref{fig:8}, we present the kinematic and orbital parameters, namely, the guiding centre radius $R_{\rm g}$, the vertical actions $J_{\rm Z}$, and the maximal height of the orbits $Z_{\rm max}$, for both the chemical thin and thick disc stars. We calculate these parameters with the {\em Galpy} \citep{Bovy2015}, utilizing the Gaia distances from \citet[][hereafter BJ21]{Bailer-Jones2021}, celestial coordinates and proper motions from Gaia EDR3 \citep{Gaia2021}. We adopt the default Milky Way potential $MWPotential2014$ for the {\em Galpy} computation. We adopt a right-hand Cartesian coordinate to calculate the space motions of our sample stars. The Sun is assumed to be located at (X, Y, Z)= ($-$8, 0, 0) kpc, and the solar motion w.r.t. the local standard of rest is ($U_{\odot}$, $V_{\odot}$, $W_{\odot}$) = (7.01, 10.13, 4.95) km s$^{-1}$ \citep{Huang2015}.

The figure illustrates that, on average, the chemical thin disc stars have smaller $J_{\rm Z}$ and $Z_{\rm max}$ than the chemical thick disc stars. However, there are substantial overlaps in the orbital parameters of individual stars between the chemical thin and thick disc. The chemical thick disc stars can have $J_{\rm Z}$ and $Z_{\rm max}$ as small as that of the thin disc stars, while the chemical thin disc stars can have $Z_{\rm max}$ as large as 1\,kpc. The results suggest that the chemically defined thin and thick disc stars have intrinsically complex kinematics, and are hard to be completely distinguished with simple criteria in kinematic parameters.

\subsection{The oldest thin disc stars}

\begin{figure*}
   \includegraphics[width=\linewidth]{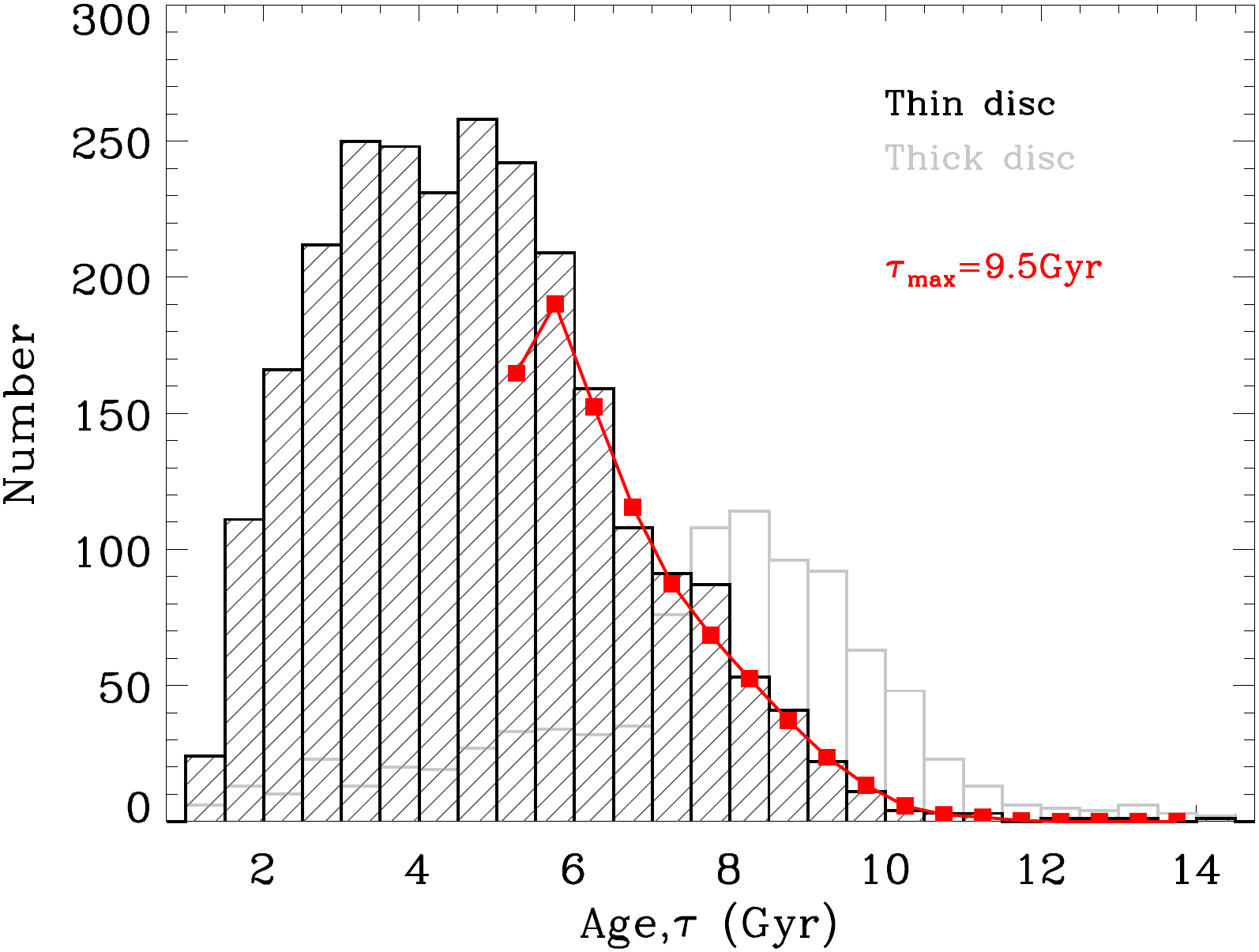}
    \caption{Age distribution histogram for the thin (black) and thick disc (grey) stars. The red line and squares represent the mock data set that best fits the the observed age histogram for determining the upper age bound of the thin disc stars (see main text). The fit suggests an upper bound of 9.5\,Gyr for the intrinsic ages of the thin disc stars.}
    \label{fig6}
\end{figure*}

Fig.~\ref{fig6} plots the stellar age distributions for the chemical thin and thick disc stars. The thin disc stars exhibit a peak age at $\sim$5\,Gyr, and there is a tail of older stars reaching beyond 8\,Gyr. The thick disc stars exhibit a peak age at $\sim$8\,Gyr, and there is a tail of younger stars reaching $\sim1$\,Gyr at the youngest end. These ``young" thick disc stars have been studied in previous work, and proved to be products of binary evolution of intrinsically old stars, while their ages are erroneously estimated to be ``young" by using single stellar evolution models \citep{Martig2015, Silva2018, Hekker2019, Sun2020, Zhang2021}. The peak age of the thick disc stars is younger than previous works with APOKASC catalogue and main-sequence turn-off and sub-giant stars, which suggest a peak age of 11\,Gyr \citep{Silva2018,Xiang2022}.
As mentioned in Sect 3.2, we suspect this is mainly because of an overestimate of effective temperatures by the LAMOST DD-Payne catalogue for high-[$\alpha$/Fe] stars. Furthermore, the photometric temperature scale of \citet{Gonz2009} is built on synthetic model spectrum of $[\alpha/{\rm Fe}]=0$ for metal-rich stars (${\rm [Fe/H]}>-0.5$) and $[\alpha/{\rm Fe}]=0.4$ for metal-poor stars (${\rm [Fe/H]}<-0.5$). These $\alpha$ mixtures are, however, not fully compatible with the real stars of  $\rm [Fe/H]\lesssim-0.3$. A better calibration of the effective temperature is required before we can reach a more realistic assessment of the age of the thick disc stars. Here we focus on the thin disc stars, for which the impact of $\alpha$ mixtures are expected to be much less significant. This is particularly true for stars with ${\rm [Fe/H]}\gtrsim-0.3$, as they have nearly solar $\alpha$ mixture.

To understand when the first thin disc stars formed, we estimate the upper bound of the age distribution of the thin disc stars with a statistical approach. The observed age distribution in Fig.~\ref{fig6} is a convolution of the intrinsic age distribution with age measurement errors. Here we intend to infer the upper age limit of the {\emph {intrinsic}} age distribution, while eliminating the impact of the measurement errors. In doing so, we assume that at the older side of the peak age ($\tau=5$\,Gyr), the intrinsic age distribution exhibits a decreasing trend that can be described with a truncated power law function
\begin{align}
    &    N(\tau) = \alpha\tau^\gamma, {\mathrm{for~} }   5<\tau<\tau_{\rm max}, \\
&    N(\tau) = 0, {\mathrm{for~} } \tau>\tau_{\rm max}.
\end{align}
The $\tau_{\rm max}$ is the maximal intrinsic age of the thin disc stars, $\gamma$ the power index, and $\alpha$ a normalization factor.

We estimate the parameters in Equations 3 and 4 with a Monte-Carlo approach. Specifically, we create mock stars for different sets of $\gamma$ and $\tau_{\rm max}$, and determine the best set of parameters by fitting the mock data to the observed age distribution, after adding a 10 per cent random error to the former to mimic the observations. We choose a set of $\gamma$ values from $-$1 to $-$5, with a step of 0.01, and a set of $\tau_{\rm max}$ from 8 to 14, with a step of 0.1\,Gyr. For each set of mock data, we compute the $\chi^2$ of the age distribution between the mock and observed data for stars of $\tau>8$\,Gyr, defined as
\begin{equation}
    \chi^2 = \Sigma\left(\frac{[N_{\rm mock}(\tau>8) - N_{\rm obs}(\tau>8)]^2}{N_{\rm mock}(\tau>8)}\right).
\end{equation}

\begin{figure*}
    \includegraphics[width=\linewidth]{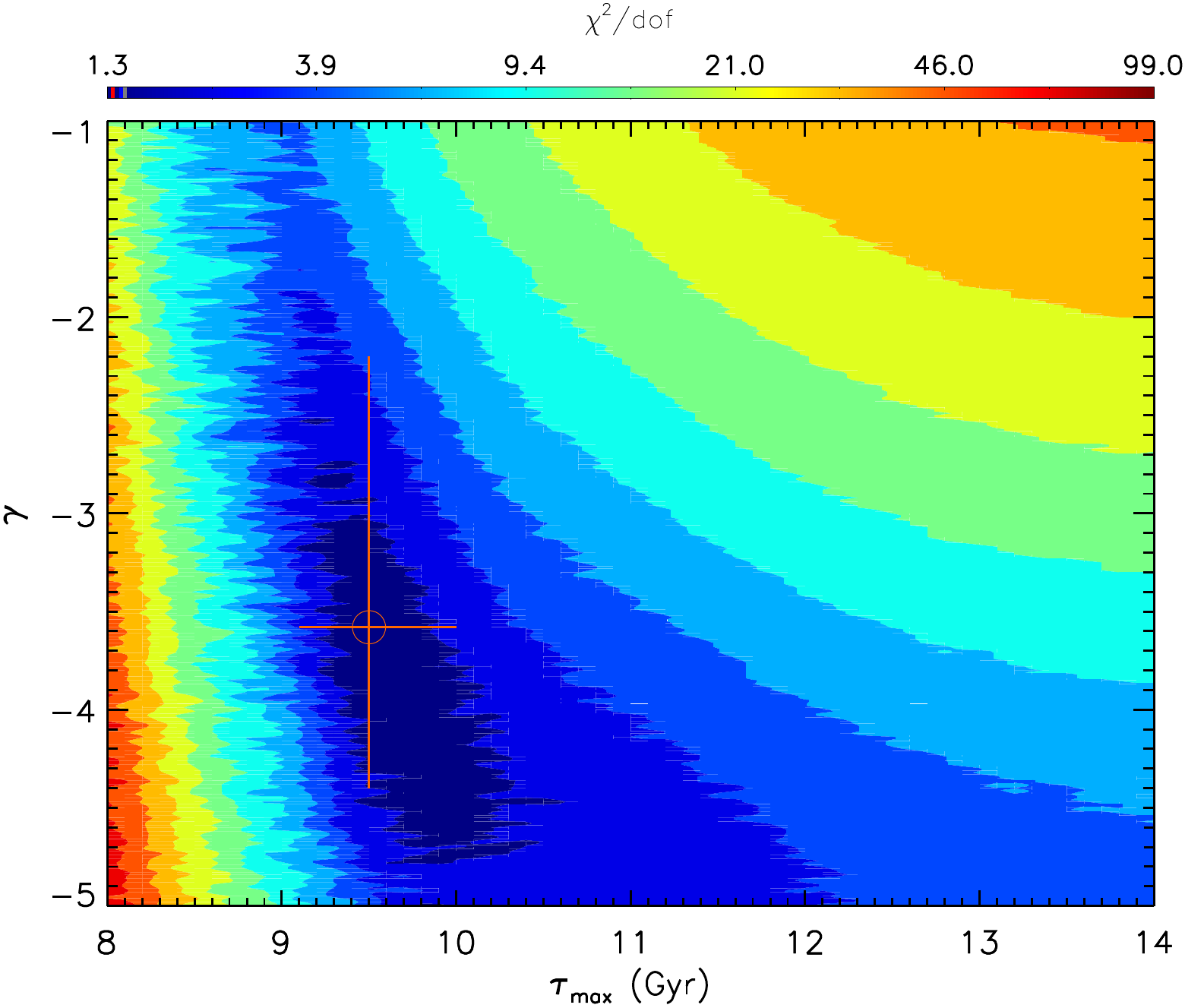}
	\caption{Color-coded contour of the $\chi^{2}$ distribution in the $\tau_{\rm max}$--$\gamma$ plane, where $\tau_{\rm max}$ is the maximal intrinsic age of the thin disc stars, $\gamma$ is the power law index of the intrinsic age distribution. The colors present the reduced $\chi^{2}$, i.e., $\chi^{2}$ divided by the degree of freedom. The orange circle with error bars represent the best-fitting parameters and their 1$\sigma$ uncertainties.}
    \label{fig7}
\end{figure*}

Fig.~\ref{fig7} shows the distribution of the reduced $\chi^2$ values (i.e., $\chi^2$/d.o.f) in the $\gamma$--$\tau_{\rm max}$ plane. The resultant parameters with the minimal $\chi^2$ value are $\gamma=-3.58$ and $\tau_{\rm max}=9.5$\,Gyr. We define the $1\sigma$ uncertainty of the parameter estimates as the deviations to parameter values that lead to a reduced $\chi^2$ value of \[\chi^2_{1\sigma}/d.o.f = \chi^2_{\rm min}/d.o.f + 1.\] The resulting maximal intrinsic age of the thin disc stars is $9.5^{+0.5}_{-0.4}$\,Gyr. The result is not very sensitive to the adopted $\gamma$, as the $1\sigma$ error bar is large ($\gamma=-3.58^{+1.38}_{-0.82}$). The age distribution of the mock data for the best-fit parameters, after adding a 10 per cent random age error, is over-plotted in Fig.~\ref{fig6}, which exhibits a good match with the observed age distribution.

\section{Discussion}

\subsection{When were the first thin disc stars formed?}
Our results suggest the oldest thin disc stars have an age of $9.5^{+0.5}_{-0.4}$\,Gyr. This value might suffer systematic uncertainties from several possible factors that we discuss in this section.

The first possible factor is the systematic uncertainties in the stellar age determinations due to imperfect stellar isochrones. Stellar isochrones are built on stellar evolution models and atmospheric models with extensive theoretical assumptions and simplifications \cite[e.g.][]{Dotter2016}. To examine possible systematics raised from imperfect stellar isochrones, we also test age determinations using other stellar isochrones, namely, the Yonsei-Yale isochrones \citep{Yi2001, Demarque2004} and the PARSEC isochrones \citep{Bressan2012}, after re-scaling the metallicity to match the LAMOST DD-Payne. We then compare the results from these models to have an estimate of possible systematic uncertainty of the age estimates. Fig.~\ref{A2} compares the age determinations from these different models. The figure illustrates that the one-to-one consistency of the results is quite good, as the dispersion of the relative age differences is only 3 per cent between the DSEP and the YY isochrones, 7 per cent between the DSEP and the PARSEC isochrones, and 8 per cent between the YY and the PARSEC isochrones. However, systematic differences are visible among these results, as the YY yields 5 per cent older age than the DSEP, while the PARSEC yields 5 per cent younger ages than the DSEP. A re-estimation of the $\tau_{\rm max}$ based on the YY stellar age estimates yields 10$\pm$0.5\,Gyr, while the value becomes  9.2$\pm$0.5\,Gyr if the PARSEC age estimates are adopted. When considering this systematic uncertainty, the first thin disc stars occur at $9.5^{+0.5(rand.)+0.5(sys.)}_{-0.4(rand.)-0.3(sys.)}$\,Gyr ago.

Secondly, as discussed in Section\,2.2, accurate effective temperatures are crucial for estimating the ages of red giant stars. A 50\,K systematic error in effective temperature will lead to more than 10 per cent uncertainty in age. The above results are obtained based on the IRFM temperature scale of \citet{Gonz2009}. We found that if the empirical temperature scale of \citet{Huang2015} that built on interferometric measurement of stellar angle size is adopted, the age estimates of our sample stars can be significantly older (Fig.~\ref{A1}), and the oldest thin disc stars would have an age of 12.4\,Gyr. However, as discussed in Section\,2.2, we suspect that adopting the \citet{Huang2015} scale for the calibration might suffer the risk of an inconsistency with the temperature of the isochrones. We therefore believe that current results based on the \citet{Gonz2009} are the realistic ones.

In addition, for characterizing the oldest thin disc stars, our results are limited by the spatial coverage of the sample, which is restricted only in a small volume (7.8\,kpc$<R<$8.6\,kpc, 0\,kpc$<Z<$1.5\,kpc). Fortunately, our sample stars should have been born at a wide range of radius between 4 and 10\,kpc, as illustrated by their guiding centre radii (Fig.~\ref{fig:8}). This gives us confidence that the current results are valid in a large volume. Undoubtedly, further studies with larger sample in a larger space volume will improve the current results.

The disc stellar age-metallicity map has also been drawn by previous studies utilizing main-sequence (turn-off) stars and subgiant stars \citep{Fuhrmann1998, Bensby2003, Nordstrom2004, Casagrande2011, Haywood2013, Bergemann2014, Xiang2017, Feuillet2019,Nissen2020, Sahlholdt2022}.
While there is no dedicated determination for the exact epoch when the \emph{first} thin disc star was formed with a method similar to the current work, most of these studies suggest the first stars have an age of roughly 8\,Gyr or older. However, a quantitative comparison requires a similar analysis to the current work using these samples. Particularly, this becomes realistic as precise ages for significant volumes of sub-giant stars have been available recently \citep{Sahlholdt2022,Xiang2022}. On the other hand, using the Galactic chemical evolution model to re-construct the double-sequence stellar distribution in [Fe/H]--[$\alpha$/Fe] plane, \citet{Spitoni2019} suggested that the Galactic disc was formed from two gas infall events (episodes), between which there is a time delay of about 4.3\,Gyr. Their results suggest that, the second gas infall episode, which corresponds to the formation of the thin disc, occurred at 9.4\,Gyr ago. This is in very good agreement with our results.

\subsection{Where were the first thin disc stars formed?}
To figure out where the first thin disc stars were born is important for building a clear picture of the early thin disc formation. However, the answer of this question is still far from well known. Based on the age--metallicity relation for a sample of $\sim1000$ main-sequence and sub-giant stars, \citet{Haywood2013} suggested that the inner thin disc ($R<10$\,kpc) experienced a different formation path with respect to the outer thin disc, and the outer disc started even earlier than the inner disc. They reach this conclusion as they found that stars in the outer disc can be older than 9 -- 10\,Gyr, while the inner disc stars are younger than 8\,Gyr. Using multi-zone chemical evolution model to re-construct the stellar distribution track in the  [Fe/H]-[$\alpha$/Fe] plane, \citet{Spitoni2021} suggest that the outer thin disc (14\,kpc$>R>$10\,kpc) started to form at 10.7\,Gyr ago, and is earlier than that of the inner disc (2\,kpc$<R<6$\,kpc), which started to form at 9\,Gyr ago.

Although our sample stars cover a small volume of 7.8\,kpc$<R<$8.6\,kpc, $Z<1.5$\,kpc (Fig.~\ref{fig4}),  it does not necessarily mean that they were born in this small part of the Galaxy. This is because the stars may have experienced migration \citep[e.g.][]{Sellwood2002,Chen2003,Roskar2008,Loebman2011,Minchev2013,Grand2016,Anders2018, Matteucci2021}. The migrators, both outward and inward ones can be identified from their different metallicities w.r.t. the in-situ stars \citep{Kordopatis2013, Frankel2018,Minchev2018, Chen2019, Zhang2021b,Wu2021}.

To precisely know the birth radii of the stars is a challenging task. Nevertheless, age and metallicity of the stars have been suggested to be effective indicators for robust estimate of their birth radii \citep{Minchev2018}. Estimates of stellar birth radii based on age and metallicity lead to qualitatively reasonable results of chemo-dynamical evolution of the thin disc stars \citep{Chen2019, Wu2021}. Similar to \citet{Chen2019} and \citet{Wu2021}, here we estimate the birth radii of our sample stars from their age and metallicity using the method of \citet{Minchev2018}.

\begin{figure*}
\includegraphics[width=\linewidth]{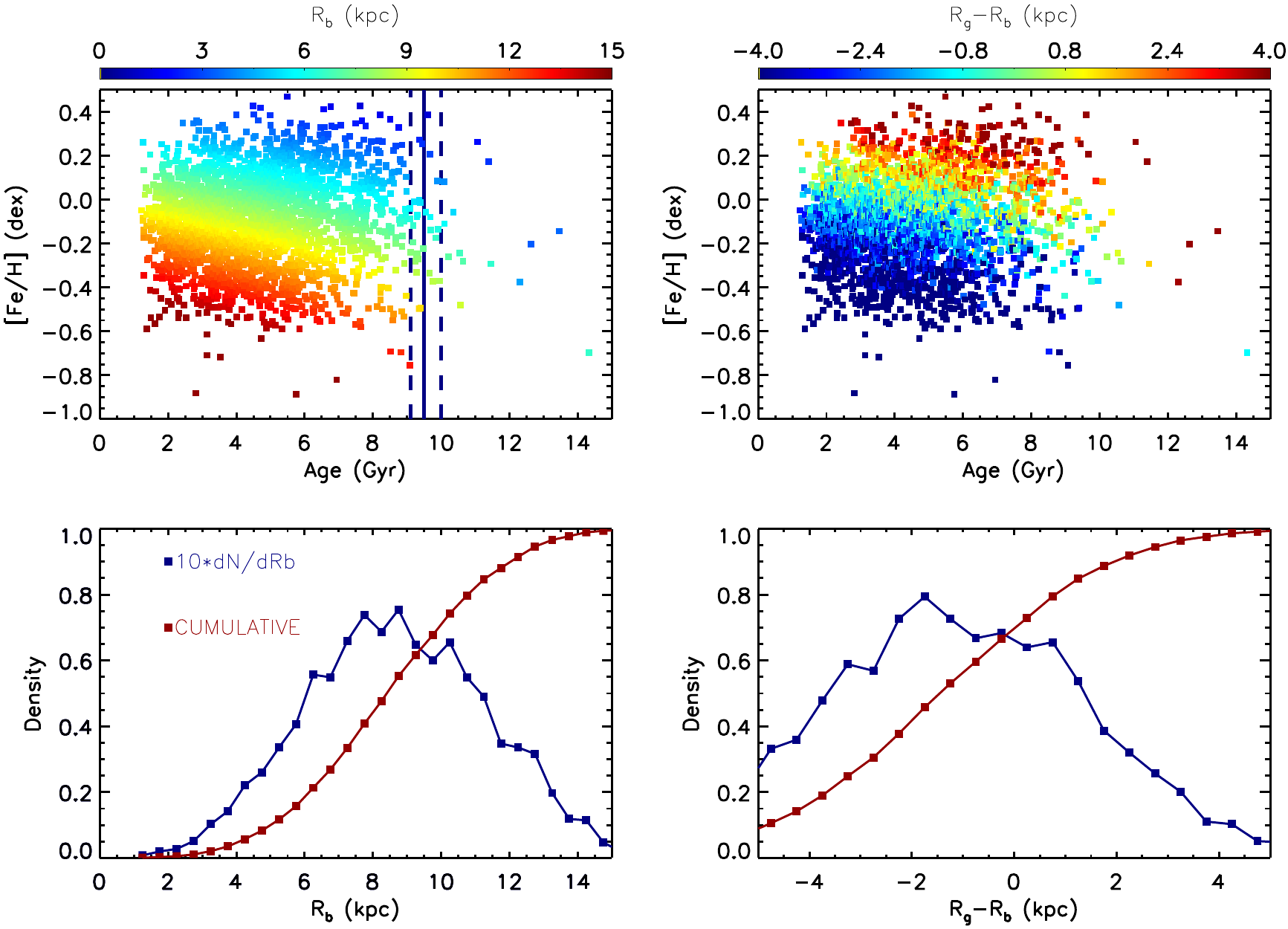}
\caption{Distributions of stellar birth radius $R_{\rm b}$ (\emph{Top-left}) as well as difference between the orbital guiding centre radius and birth radius $R_{\rm g}-R_{\rm b}$ (\emph{Top-right}) for the chemical thin disc stars in the age-[Fe/H] plane. The vertical solid line represents a constant age of 9.5\,Gyr, the estimated upper \textbf{bound} of the intrinsic ages of the thin disc stars, and the dotted lines represent uncertainties of bound(\emph{Top-left}). The differential and the cumulative distribution
of the stellar birth radius $R_{\rm b}$ and of $R_{\rm g}-R_{\rm b}$ are shown in the left- and right-bottom panels, respectively.} \label{fig10}
\end{figure*}

Fig.~\ref{fig10} shows the distribution of the birth radii $R_{\rm b}$, as well as the difference between the birth radii and guiding-centre $R_{\rm g}$ of the chemical thin disc stars in the age -- metallicity plane. Here, the difference between $R_{\rm g}$ and $R_{\rm b}$ is a direct measure of the stellar redistribution due to radial migration. The figure illustrates that the birth radii of our sample stars cover a large range, from $\sim$3\,kpc for the old metal-rich stars to $\gtrsim12$\,kpc for the young, metal-poor stars. The most metal-rich stars (${\rm [Fe/H]}\gtrsim0.2$) were migrated outward from the very inner disc by more than 4\,kpc, while the metal-poor thin disc stars (${\rm [Fe/H]}\lesssim-0.4$) were migrated inward from the outer disc by more than 4\,kpc. Typical uncertainty of the birth radius is about 1\,kpc \citep{Chen2019,Wu2021}, while the uncertainty in the guiding centre radius is negligible. About 73 per cent of our sample stars exhibit difference in their $R_{\rm g}$ and $R_{\rm b}$ larger than 1\,kpc, and about 16 per cent of them is even larger than 4\,kpc, suggesting the radial migration is an important effect for explaining the presence of old metal-rich and young metal-poor stars in our sample.

We note that the detailed role of radial migration on the formation of the disc is still in debated (see more in \citet{Matteucci2021}). Our results suggest that the old metal-rich and young metal-poor stars may have experienced strong migration to reach their current position, which is consistent with previous studies \citep[e.g.][]{Minchev2013,Kordopatis2015,Wu2018,Chen2019,Chen2020,Wu2021}. However, to assess the role of radial migration on the formation of the Galactic disc in a larger context is beyond the scope of our discussion here, as we focus on characterizing where the first stars were formed.

 Fig.~\ref{fig10} illustrates that the oldest thin disc stars ($\tau\simeq9$\,Gyr) can have ${\rm [Fe/H]}$ values ranging from $-0.6$ to $0.4$, and may have been born at any Galactocentric radius from $R_{\rm b}\simeq3$\,kpc to $R_{\rm b}\sim10$\,kpc. This implies that the inner and outer Galactic thin disc might have started to form stars at approximately the same time. Equivalently, for all Galactocentric radii of $R_{\rm b}\lesssim10$\,kpc, the oldest stars can have an age of $\gtrsim$9\,Gyr. This is slightly different to some previous work that  suggest the outer disc formed earlier as mentioned above \citep[e.g.][]{Spitoni2021}. This difference may be due to the fact that we adopt $R_{\rm b}$ to define the inner and outer disc, while in the literature, the current Galactocentric radius $R$ is commonly adopted. Here we caution that because the disc stars may have experienced significant radial migration, we prefer $R_{\rm b}$ rather than $R$ to define the `inner' and `outer' disc.

It is also worth mentioning that our result does not necessarily conflict with the popular inside-out disc formation scenario. The inside-out disc formation scenario refers to the dependence of the time for the {\em peak} star formation rate as a function of Galactocentric radius. A negative radial age gradient has been observed in previous work \citep[e.g.][]{Martig2016-thick-disc, Xiang2018, Frankel2019, Wu2019}, which is consistent with the inside-out scenario. However, here we focus on the birth radii of the first thin disc stars, i.e., stars with the oldest ages, rather than the mean age that corresponds to the peak star formation rate.

\subsection{Co-formation with the thick disc?}

Recent studies have argued that the thin and thick discs are co-formed \citep[e.g.][]{Silva2021,Gent2022}. This conclusion is reached mainly by the fact that some thin disc stars can be as old as the thick disc stars. As discussed above, age estimates for thick disc stars in the current work suffer severe systematic uncertainties due to temperature calibration, so that it is not possible for us to make a direct comparison of ages between the thin and thick disc stars. However, as shown by \citet{Xiang2022}, the thick disc has experienced a long star-formation episode started at 13\,Gyr ago and quenched at 7--8\,Gyr ago, with a peak of star-formation rate at 11\,Gyr ago. This means that the formation of the first thin disc stars (9.5\,Gyr) is unavoidably overlapped with the thick disc formation. To avoid ambiguity, we should note that the epochs for peak star-formation rate are very different between the thin and thick discs, with most thin disc stars formed at 4-5\,Gyr (Fig.~\ref{fig6}), which is much younger than the thick disc.

\section{Conclusion}

The formation of the extended thin disc has been the most spectacular event of our Galaxy in the past $\sim$\,8\,Gyr. Unravelling the start of this event requires precise and accurate stellar ages, which we
aim to provide in the current work.

We have investigated the asteroseismic age determination of 5306 red giant branch stars using \kepler{} and LAMOST data, with a thorough examination of how age determination is affected by choice of different temperature scales and stellar models. The typical age precision of these sample stars is better than 12 per cent, thus allowing a robust characterization for the age of the oldest stars in the Galactic thin disc. Our conclusions are as follows.
\begin{itemize}
    \item We found that the temperature scale has a significant impact on age determination of RGB stars, as 50\,K difference in effective temperature may cause larger than 10 per cent systematic error in the age estimates. We have also characterized systematic uncertainty in the asteroseismic age estimates from stellar models by comparing the ages derived from different sets of isochrones that are publicly available, and found a typical error of about 0.5\,Gyr.
    \item We found that the oldest thin disc stars have an age of $9.5^{+0.5(rand.)+0.5(sys.)}_{-0.4(rand.)-0.3(sys.)}$\,Gyr. These stars can have ${\rm [Fe/H]}$ values ranging from $-$0.6 to 0.4, and may be born at all Galactocentric radii from $\sim$3\,kpc to $\sim10$\,kpc. This implies the inner and outer thin disc have begun to form at the same time, rather than displaying a delay. At this epoch, the Galactic thick disc is still forming stars, suggesting there is a time overlap for the thin and thick disc formation, although the epochs for the peak star-formation rate is different between the two discs.
\end{itemize}

\section*{Acknowledgements}

It is a pleasure to thank the anonymous referee for helpful
suggestions that have significantly improved the presentation of the manuscript. This work is supported by the National Natural Science Foundation of China (NSFC) under grant No.11988101, 11903044, 11973049, the Joint Research Fund in Astronomy (U2031203), National Key R\&D Program of China No. 2019YFA0405502. We also acknowledge from the China Manned Spaced Project with NO.CMS-CSST-2021-B05. M.X. acknowledges financial support from NSFC Grant No.2022000083. and National Key R\&D Program of China Grant No. 2022YFF0504200.

Guoshoujing Telescope (the Large Sky Area Multi-Object Fiber Spectroscopic Telescope LAMOST)
is a National Major Scientific Project built by the Chinese Academy of Sciences.
Funding for the project has been provided by the National Development and Reform Commission.
LAMOST is operated and managed by the National Astronomical Observatories, Chinese Academy of Sciences.

This work has also made use of data from the European Space Agency (ESA) mission
Gaia \footnote{https://www.cosmos.esa.int/gaia}, processed
by the Gaia Data Processing and Analysis Consortium
(DPAC, \footnote{https://www.cosmos.esa.int/web/gaia/dpac/consortium}). Funding for the DPAC has been provided
by national institutions, in particular the institutions
participating in the Gaia Multilateral Agreement.

\section*{Data Availability}
The data underlying this article will be shared on reasonable request to the corresponding author.




\bibliographystyle{mnras}
\bibliography{timing_clean} 




\appendix

\section{\textbf{Impact of temperature scale on age determination}}

\begin{figure*}
   \includegraphics[width=\linewidth]{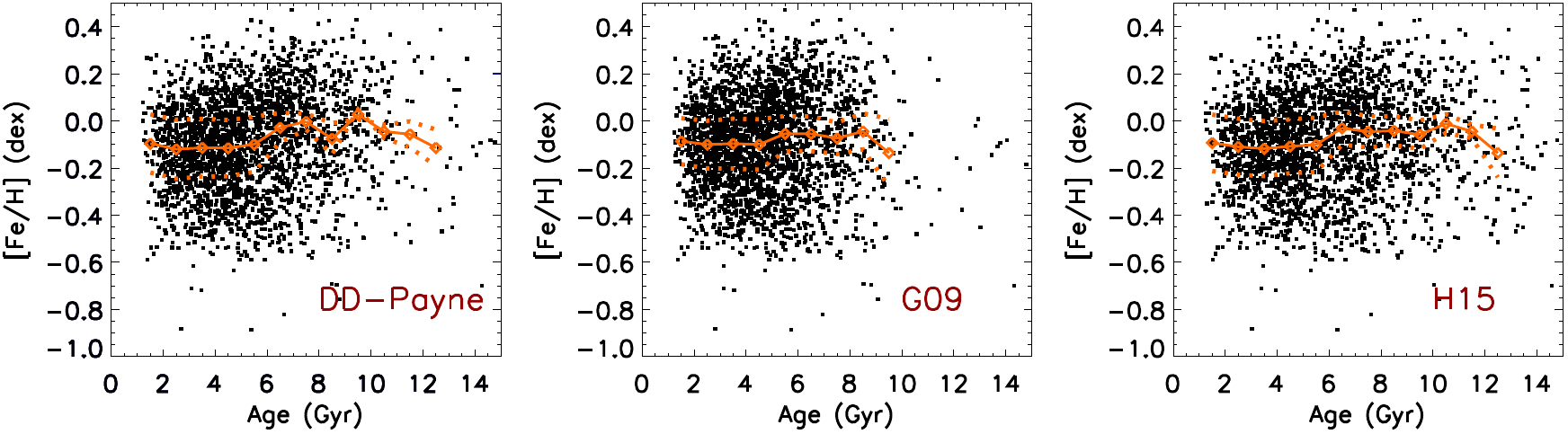}
    \caption{The thin disc's stellar age-[Fe/H] relation for age estimates derived using effective temperature of different scales. The \emph{left} panel shows ages estimated directly using the LAMOST DD-Payne temperature, without any calibrations. The \emph{middle} panel shows ages estimated after calibrating the LAMOST DD-Payne temperature to the \citet{Gonz2009} IRFM scale. The \emph{right} panel shows ages estimated after calibrating the LAMOST DD-Payne temperature to the \citet{Huang2015} scale. Clear differences in the ages are seen for the old stars among these panels (see text).}

    \label{A1}
\end{figure*}

As discussed in Section\,2.2, accurate temperatures are crucial to estimate ages of red giant stars. This is because in the RGB phase stellar isochrones of different ages exhibit close effective temperatures, any small systematic error in the effective temperature could cause a large deviation in the age determination. We plot the age--[Fe/${\rm H}$] relation of our sample stars resulting from different temperature scales in Fig.~\ref{A1}. In general, the figure shows a broad [Fe/H] distribution at all ages from 1 to 13\,Gyr. On the other hand, the figure shows that both no calibration to the LAMOST DD-Payne temperature and calibrating it to the \citet{Huang2015} scale will lead to significantly more old stars than results using temperature calibrating to the \citet{Gonz2009} scale. Particularly, the scale of \citet{Huang2015} will lead to much more old stars than the other two cases. These will impact the estimated upper age bound of the Galactic thin disc. Specifically, we found that there is no correction in the LAMOST DD-Payne temperature, the oldest thin disc stars will be $\sim11.2$\,Gyr, while if the \citet{Huang2015} scale is adopted, the oldest thin disc stars will be 12.4\,Gyr. However, as discussed in the main text, we tend to believe that the \citet{Gonz2009} scale gives the most reasonable results.

\section{\textbf{Impact of stellar models on age determination}}
\begin{figure*}
    \includegraphics[width=\linewidth]{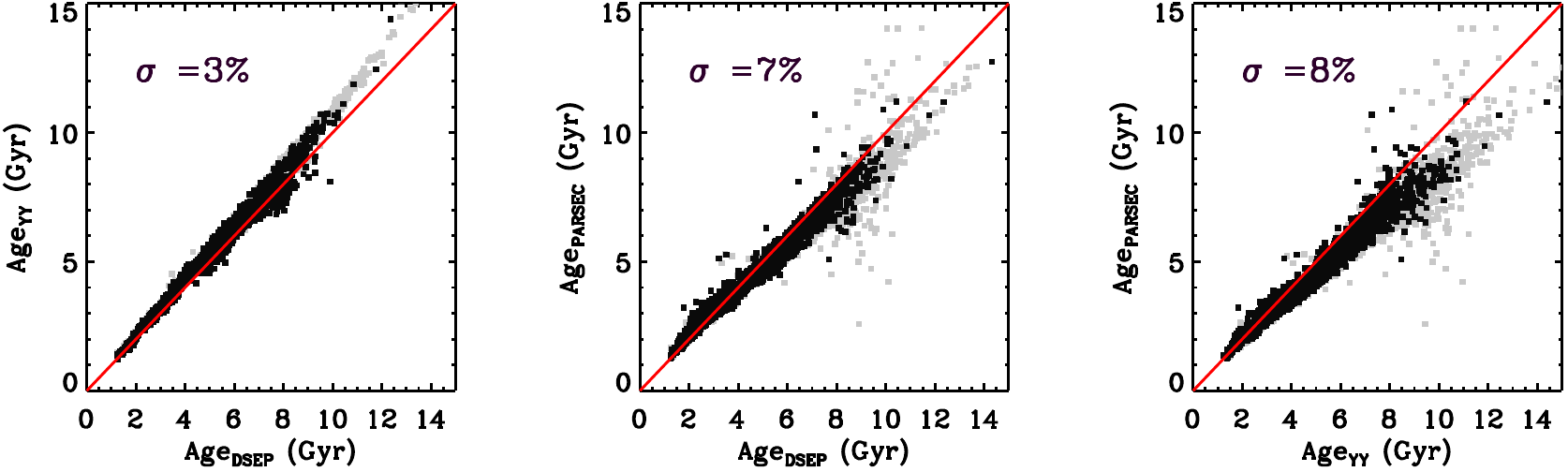}
    \caption{Comparison of age estimates using different sets of stellar evolutionary isochrones, namely, the Dartmouth stellar evolution project \citep[DSEP;][]{Dotter2008}, the Yonsi-Yale \citep[YY;][]{Demarque2004}, and the PARSEC \citep{Bressan2012} isochrones. In all panels, black dots represent chemical thin disc stars, while grey dots represents chemical thick disc stars. The red line represents the 1:1 line. Dispersion of the relative age difference is marked in the figure.}

    \label{A2}
\end{figure*}

We have also examined possible systematic uncertainties in the age estimates caused by imperfect stellar isochrones. We do this by comparing results from different stellar evolutionary databases, namely, the Dartmouth stellar evolution project \citep[DSEP;][]{Dotter2008}, the Yonsi-Yale \citep[YY;][]{Demarque2004}, and the PARSEC \citep{Bressan2012} isochrones. A comparison of the resulting age determinations is shown in Fig.~\ref{A2}. The figure illustrates that the one to one relation of the results are quite good, as the dispersion of the relative age differences is only 3 per cent between the DSEP and the YY isochrones, 7 per cent between the DSEP and the PARSEC isochrones, and 8 per cent between the YY and the PARSEC isochrones. However, systematic differences are visible among these results, as the YY yields 5 per cent older age than the DSEP, while the PARSEC yields 5 per cent younger ages than the DSEP. A re-estimation of the $\tau_{\rm max}$ based on the YY stellar age estimates yields 10$\pm$0.5\,Gyr, while the value becomes 9.2$\pm$0.5\,Gyr if the PARSEC isochrones are adopted. This makes us conclude that the oldest thin disc stars have an age of $9.5^{+0.5(rand.)+0.5(sys.)}_{-0.4(rand.)-0.3(sys.)}$.

\bsp	
\label{lastpage}
\end{document}